\documentstyle[12pt]{article}

\begin{document}
\title{\bf {Casimir Effect for Spherical Shell in\\ de Sitter
Space}}
\author{
M.R. Setare $^1$ \footnote{E-mail: Mreza@physics.sharif.ac.ir}
\\  R. Mansouri $^{1,2}$\footnote{E-mail: mansouri@sina.sharif.ac.ir}  \\
 {$^1$ Department of Physics, Sharif University of
Technology, Tehran, Iran}\\ and \\ {$^{1,2}$ Institute for
Theoretical Physics and Mathematics, Tehran, Iran}}
\date{\small{\today}}
 \maketitle
\begin{abstract}
The Casimir stress on a spherical shell in de Sitter background
for massless scalar field satisfying Dirichlet boundary conditions
on the shell is calculated. The metric is written in conformally
flat form. Although the metric is time dependent, no particles
are created. The Casimir stress is calculated for inside and
outside of the shell with different backgrounds corresponding to
different cosmological constants. The detail dynamics of the
bubble depends on different parameter of the model. Specifically,
bubbles with true vacuum  inside expands if the difference in the
vacuum energies is small, otherwise it collapses.
 \end{abstract}
% \begin{document}
\newpage
% \vspace*{10mm}

 \section{Introduction}
   The Casimir effect is one of the most interesting manifestations
  of nontrivial properties of the vacuum state in quantum field
  theory[1,2]. Since its first prediction by
  Casimir in 1948\cite{Casimir} this effect has been investigated for
  different fields having different boundary geometries[4-7]. The
  Casimir effect can be viewed as a polarization of
  vacuum by boundary conditions or geometry. Therefore, vacuum
  polarization induced by an gravitational field is also considered as
  Casimir effect.\\
  A new element which has recently been taking into account is to bring
  dynami
  cs regarding the boundary conditions or the
  geometry into this effect. This dynamical Casimir effect have
  been studied by different authors[8-10]. In the static case
  perturbation of the quantum state induces vacuum energy
  and stress but no particle. In contrast, in the vacuum perturbed by
  dynamical external constrains particles are created.
  For example, a scalar massless field propagating between two infinite
  parallel plates moving with constant relative velocity creates
  particles at the expense of the Casimir energy, or even the motion
  of a single reflecting boundary can cause such an effect.
  \cite{{Birrell},{Hacyan}}. Creation of particles by time-dependent gravitational
  field is another example of such dynamical effects.
  Taking different possible dynamical effects into account, one may wonder
  how Casimir effect may correct our view of the early universe. It has been
  shown, e.g., in\cite{Ant} that a closed Robertson-Walker space-time in
  which the only contribution to the  stress tensor comes from Casimir energy
   of a scalar field is excluded. In inflationary models, where the
   dynamics of bubbles may play a major role, this dynamical
   Casimir effect has not yet been taken into account.\\
   Casimir effect for spherical shells in the presence of the
   electromagnetic fields has been calculated several years ago[13,14]. A
   recent simplifying account of it for the cases of electromagnetic and
   scalar field with both Dirichelt and  Neumann boundary conditions on sphere
   is given in\cite{Nester}. The dependence of Casimir energy for scalar
   and electromagnetic fields with Dirichlet boundary conditions
  in the presence of a spherical shell is discussed in[12,16]. It has been
  shown that the Casimir energy in even space dimensions, in contrast to the
  case of odd dimensions, is divergent. Spinor fields are considered
   in\cite{Cog}. Robin's boundary conditions have been studied in\cite{Cog}, where
   interior and exterior regions are treated separately. It is shown explicitly
   that although the Casimir energy for interior and exterior of the shell are both
   divergent irrespective of the number of space dimensions, the total
   Casimir energy of the shell remains finite for the case of odd space
    dimensions. Of some interest are cases where the field is
    confined to the inside of a spherical shell. This is sometimes called
    as the bag boundary condition. The application of Casimir effect to the
    bag model is considered for the case of massive scalar field
    \cite{Bord} and the Dirac field \cite{Eli2}. The
    renormalization procedure in the above cases is of interest
    and we use it for the cases of interest to us.\\
    There are few examples of Casimir effect in curved space-time.
    Casimir effect for spherical boundary in curved space-time is
    considered in[21, 22], where the Casimir energy for
  half of $s_{3}$ and $s_{2}$ with Dirichlet and Neumann boundary
  conditions for massless conformal scalar field is
  calculated analytically using all the existing methods.
  Casimir effect in the presence of a general relativistic domain
  wall is considered in \cite{Set} and a study on the relation
  between trace anomaly and the Casimir effect can be found in
  \cite{set1}.\\
   Non of these cases imply dynamical effects.
  Our aim is to consider the dynamical Casimir effect on a
   spherical shell having different vacuums inside and outside
   representing a bubble in early universe with false/true vacuum
   inside/outside. Section two is devoted to the Casimir effect for a
   spherical shell with Dirichelt boundary conditions. In section
   three we calculate the stress on a spherical shell having constant
   comoving radius in a de Sitter space. The case of different de
   Sitter vacuums inside and outside of the shell, using the renormalization
   method of MIT bag model, is considered in section four. In last
   section we conclude and summarize the results.

  \section{Scalar Casimir effect for a sphere in flat space-time}

  Consider the Casimir force due to fluctuations of a free massless
  scalar field satisfying Dirichlet boundary conditions on a spherical
  shell in Minkowski space-time \cite{Milton}. The two-point Green's
  function $G(x,t;x',t')$ is defined as the vacuum expectation
  value of the time-ordered product of two fields
  \begin{equation}
  G(x,t;x',t')\equiv-\imath<0|T\Phi(x,t)\Phi(x',t')|0>.
  \end{equation}
  It has to satisfy the Dirichlet boundary conditions on the
  shell:
  \begin{equation}
  G(x,t;x',t')|_{|x|=a}=0,
  \end{equation}
  where $a$ is radius of the spherical shell. The stress-energy tensor
  $T^{\mu\nu}(x,t)$ is given by
  \begin{equation}
  T^{\mu\nu}(x,t)\equiv\partial^{\mu} \Phi(x,t)\partial^{\nu} \Phi(x,t)-
  \frac{1}{2}\eta^{\mu\nu} \partial_{\lambda} \Phi(x,t)\partial^{\lambda}
  \Phi(x,t).
  \end{equation}
  The radial Casimir force per unit area $\frac{F}{A}$ on the
  sphere, called Casimir stress, is obtained from the radial-radial component of the vacuum
  expectation value of the stress-energy tensor:
  \begin{equation}
  \frac{F}{A}=\langle0|T^{rr}_{in}-T^{rr}_{out}|0\rangle|_{r=a}.
  \end{equation}
  Taking into account the relation (1) between the vacuum expectation value
  of the stress-energy tensor $T^{\mu\nu}(x,t)$ and the Green's
  function at equal times $ G(x,t;x',t)$ we obtain
  \begin{equation}
  \frac{F}{A}=\frac{\imath}{2}[\frac{\partial}{\partial r}\frac{\partial}{\partial
  r'}G(x,t;x',t)_{in}-\frac{\partial}{\partial r}\frac{\partial}{\partial
  r'}G(x,t;x',t)_{out}]|_{x=x',|x|=a}.
  \end{equation}

  \section{Scalar Casimir effect for a sphere in de Sitter space}
 Consider now a massless scalar field in de Sitter space-time. To
make the maximum use of the flat space calculation we use the de
Sitter metric in conformally flat form:
  \begin{equation}
  ds^{2}=\frac{\alpha^{2}}{\eta^{2}}[d\eta^{2}-\sum_{\imath=1}^{3}(dx^{\imath})^{2}],
  \end{equation}
  where $\eta$, is the conformal time:
  \begin{equation}
  -\infty\langle  \eta  \langle 0.
  \end{equation}
  Under the conformal transformation in four dimensions the scalar
  field $\Phi(x,t)$ is given by
    \begin{equation}
  \bar\Phi(x,\eta)=\Omega^{-1}(x,\eta)\Phi(x,\eta).
  \end{equation}
  With the conformal factor given by
  \begin{equation}
  \Omega(\eta)=\frac{\alpha}{\eta}.
  \end{equation}
 And assuming a canonical quantization of the scalar field, and using
  the creations and annihilations operators $a_{k}^{\dagger}$ and
  $a_{k}$, the scalar field $\Phi(x,\eta)$ is then given by
\begin{equation}
  \Phi(x,\eta)=\Omega(\eta)\sum_{k}[a_{k}
  \bar u_{k}(\eta,x)+a_{k}^{\dagger}\bar u_{k}^{\ast}(\eta,x)]
  \end{equation}
  The vacuum states associated with the modes $\bar u_{k}$ defined by
  $a_{k}|\bar 0\rangle=0 $, are called conformal vacuum. For the massless
  scalar field we are considering, the Green's function $\bar G$
  associated to the conformal vacuum$|\bar 0\rangle$ is given by the
  flat Feynman Greens function times a conformal factor\cite{{Birrell},{Hartle}}.
  Given the flat space Green's function(1), we obtain
  \begin{equation}
 \bar G=-\imath\langle\bar{0}|T \bar \Phi(x,\eta)\bar
 \Phi(x',\eta^{'})|\bar{0}\rangle=\Omega^{-1}(\eta)\Omega^{-1}(\eta^{'})G.
 \end{equation}
Therefore, the stress(5) is given by
\begin{equation}
(\frac{\bar F}{A})_{in}=\frac{\imath}{2}[\frac{\partial}{\partial
r }\frac{\partial}{\partial r'}\bar
G(x,\eta;x',\eta)_{in}]|_{x=x',|x|=a}=\frac{\eta^{2}}{\alpha^{2}}(\frac{F}{A})_{in},
\end{equation}
and similarly
\begin{equation}
(\frac{\bar
F}{A})_{out}=\frac{\eta^{2}}{\alpha^{2}}(\frac{F}{A})_{out}.
\end{equation}
Finally, taking the definition (4), we obtain for the total stress
on the sphere
\begin{equation}
(\frac{\bar F}{A})=\frac{\eta^{2}}{\alpha^{2}} \frac{F}{A}.
\end{equation}
Now we consider the pure effect of vacuum polarization due to the
gravitational field without any boundary conditions. The
renormalized stress tensor for massless scalar field in de Sitter
space is given by\cite{{Birrell}, {Dowker}}:
\begin{equation}
\langle T^{\nu}_{\mu}\rangle=\frac{1}{960 \pi ^{2}\alpha
^{4}}\delta^{\nu}_{\mu}.
\end{equation}
The corresponding effective pressure is then
\begin{equation}
P=-\langle T^{1}_{1}\rangle=-\langle
T^{r}_{r}\rangle=-\frac{1}{960 \pi^{2}\alpha^{4}},
\end{equation}
valid for both in and out side of the sphere. Hence, the effective
force on the sphere is zero.\\
The particle creation in such cases is a delicate problem. The
metric (6) has an apparent time dependence. On the other side, it
is conforml to Minkowski space and also the scalar field is
massless and conformally coupled to de Sitter background.
Therefore, we may not expect any particle production. Particle
production takes place only when the conformal symmetry is broken
by the presence of mass, which provides a lengh scale for the
 theory\cite{Birrell}. To see this explicitly, we calculate the
 corresponding Bogolubov coefficients. Free massless scalar
 field $\Phi(x,\eta)$ in Minkowski space-time satisfies the Klein-Gordon
 equation
 \begin{equation}
 (\frac{\partial^{2}}{\partial\eta^{2}}-\nabla^{2})\Phi(x,\eta)=0.
\end{equation}
To solve this equation we introduce polar coordinates and seek a
solution that has cylindrical symmetry, we seek a solution that is
a function only of two variables $r=|x|$ and $\theta$,the angle
between $x$ and$x'$ so that $x.x'=rr'\cos\theta$ \cite{Milton}. In
terms of these polar variables(17) becomes
\begin{equation}
[\frac{\partial^{2}}{\partial\eta^{2}}-(\frac{\partial^{2}}{\partial
r^{2} }+\frac{2\partial}{r\partial r}+\frac{1}{\sin\theta
r^{2}}\frac{\partial}{\partial\theta}\frac{\sin\theta\partial}{\partial\theta})]\Phi(r,\theta,\eta)=0
\end{equation}
We can solve (18) using the method of separation of variables. Let
\begin{equation}
\Phi(r,\theta,\eta)=A(r)B(\theta)T(\eta),
\end{equation}
and
\begin{equation}
T(\eta)=\exp^{-\imath\omega\eta}.
\end{equation}
The scalar field $\Phi(r,\theta,\eta)$in de Sitter space satisfies
\begin{equation}
(\Box+\xi R)\bar\Phi(r,\theta,\eta)=0
\end{equation}
where $\Box$ is the Laplace-Beltrami operator for the de Sitter
metric, and $\xi$ is the coupling constant. For conformally
coupled field in four dimension $\xi=\frac{1}{6}$, and R , the
Ricci scalar curvature, is given by
\begin{equation}
R=12\alpha^{-2}.
\end{equation}
 Now, the Bogolubov transformation, given by
\begin{equation}
u_{k}^{in}(r,\theta,\eta)=\alpha_{k}u_{k}^{out}(r,\theta,\eta)+\beta_{k}u_{-k}^{\ast
out}(r,\theta,\eta),
\end{equation}
defines the Bogoloubov coefficients $\alpha_{k}$ and $\beta_{k}$.
Here $"^{in}"$ and $"^{out}"$ corresponds to $(\eta\rightarrow
-\infty)$ and$(\eta\rightarrow T<0)$ respectively. Taking into
account the separation of variables (19) we obtain from (23):
\begin{equation}
T_{k}^{in}(\eta)=\alpha_{k}T_{k}^{out}(\eta)+\beta_{k}T_{k}^{\star
out }(\eta).
\end{equation}
 Due to(8) and(20) we may write
\begin{equation}
\eta\exp({-\imath\omega\eta})=\alpha_{k}\eta\exp({-\imath\omega\eta})+
\beta_{k}\eta\exp({\imath\omega\eta}).
\end{equation}
Therefore
\begin{equation}
\alpha_{k}=1  \hspace{2cm} \beta_{k}=0.
\end{equation}
 But the expectation value of the number operator
$N_{k}=a_{k}^{\dag}a_{k}$ for the number of $\bar u_{k}$-mode
particles in the state $|\bar 0\rangle$ is given by\cite{Birrell}
\begin{equation}
\langle \bar 0|N_{k}|\bar 0\rangle=\sum_{j}|\beta_{jk}|^{2},
\end{equation}
which is zero in our case. Therefore there is no particle
production in our case.

 \section{Spherical shell with different vacuum inside and
    outside}

Now, assume there are different vacuums in- and out-side,
corresponding to $\alpha_{in}$ and $\alpha_{out}$ for the metric
(6). It is then more suitable to use the following relation for
the stress on the shell\cite{Schw}:
\begin{equation}
\frac{F}{A}=<T_{rr}>_{in}-<T_{rr}>_{out}=\frac{-1}{4\pi
a^{2}}\frac{\partial E}{\partial a},
\end{equation}
where $E$ is Casimir energy due to boundary conditions. The
Casimir energy $E$ is the sum of Casimir energies $E_{in}$
and$E_{out}$ for inside and outside of the shell.
 As
described in introduction, Casimir energies in-side and out-side
of the shell are divergent individually. In flat space when we
calculate the total Casimir energy, we add interior and exterior
energies to each other. Now divergent parts will cancel each other
out, when interior and exterior background are the same, like the
case mentioned in last section,we get the above result again.

 In flat space-time for massless scalar field with Dirichlet
boundary conditions the Casimir energy in- and out-side of a
spherical shell is given by\cite{Cog}
\begin{equation}
E_{in}=\frac{1}{2a}(0.008873+\frac{0.001010}{\epsilon})
\hspace{2cm}
E_{out}=\frac{-1}{2a}(0.003234+\frac{0.001010}{\epsilon}).
\end{equation}
Each of the energies for in-and out-side of the shell is
divergent, and cutoff dependent. But the Casimir energy, which is
the sum of $E_{in}$ and $E_{out}$ is independent of the cutoff
$\epsilon$ and finite. In the case of different in- and out-side
backgrounds, the boundary part of the total Casimir energy is
calculated to be
\begin{equation}
\bar
E_{in}=\frac{\eta^{2}}{2a\alpha_{in}^{2}}(c_{1}+\frac{c_{1'}}{\epsilon})
\hspace{1cm} \bar E_{out}=\frac{\eta^{2}}{2a
\alpha_{out}^{2}}(c_{2}-\frac{c_{1'}}{\epsilon}),
\end{equation}
where, $c_{1} = 0.008873$, $c_{2} = -0.003234$, $c_{1'}=0.001010$.
In this case, we have
\begin{equation}
\bar E=\bar E_{in}+\bar
E_{out}=\frac{\eta^{2}}{2a}(\frac{c_{1}}{\alpha_{in}^{2}}+
\frac{c_{2}}{\alpha_{out}^{2}})+ \frac{\eta^{2}c_{1'}}{2a
\epsilon}(\frac{1}{\alpha_{in}^{2}}-\frac{1}{\alpha_{out}^{2}}).
\end{equation}
Therefore the Casimir energy for this general case becomes cutoff
dependent and divergent. To renormalize  the Casimir energy $\bar
E$, we use a procedure similar to that of the bag model[18-19].
The classical energy of a spherical shell, or bubble, immersed in
a cosmological background, as we are considering, maybe written as
\begin{equation}
E_{(class)} = pa^3 + \sigma a^2 + Fa + K + \frac{h}{a},
\end{equation}
where the meaning of the terms proportional to $P$ and $\sigma$
is obvious. The third and forth terms on the right hand side of
the above equation corresponds to the curvature and cosmological
term respectively. The last term is considered as non-vanishing
because of the intuition obtained from the calculation of the
Casimir effect in the last section. There we have seen that the
Casimir energy on each side of the bubble is proportional to
$\frac{1}{a}$. Terms proportional to other powers of $a$ is
therefore not expected. Now, in our case the total Casimir energy
is divergent and therefore we have to renormalize the parameter
$h$. The total energy of the shell maybe written as
\begin{equation}
\tilde{E_{in}}=\bar{E_{in}}+E_{(class)} \hspace{2cm}
\tilde{E_{out}}=\bar{E_{out}}+E_{(class)}.
\end{equation}
The renormalization can be achieved now by shifting the parameter
h of $E_{(class)}$ by an amount which cancels the divergent
contribution. For inside and outside we have
\begin{equation}
h\rightarrow h-\frac{\eta^{2}c_{1'}}{2a\epsilon\alpha_{in}^{2}}
\hspace{1cm} h\rightarrow
h+\frac{\eta^{2}c_{1'}}{2a\epsilon\alpha_{out}^{2}}.
\end{equation}
We finally obtain for the total zero point energy of our system:
\begin{equation}
\bar{E}=\frac{\eta^{2}}{2a}(\frac{c_{1}}{\alpha_{in}^{2}}+\frac{c_{2}}{\alpha_{out}^{2}}).
\end{equation}
In contrast to the Minkowski- and de Sitter-space the Casimir
energy can now be positive or negative depending on the value of
$\alpha$ in- and out-side of the bubble. The stress on the shell
due to boundary conditions is then obtained(28):
\begin{equation}
\frac{\bar{F}}{A} = \frac{-1}{4\pi
a^{2}}\frac{\partial\bar{E}}{\partial a}=\frac{\eta^{2}}{8\pi
a^{4}}(\frac{c_{1}}{\alpha_{in}^{2}}+\frac{c_{2}}{\alpha_{out}^{2}}).
\end{equation}
As expected, one obtain the previous result (14) for the case
$\alpha_{in}=\alpha_{out}$. Now, the effective pressure created by
gravitational part(16), is different for different part of
space-time:
\begin{equation}
P_{in}=-<T_{r}^{r}>_{in}=\frac{-1}{960\pi^{2}\alpha_{in}^{4}}
\hspace{2cm}
P_{out}=-<T_{r}^{r}>=\frac{-1}{960\pi^{2}\alpha_{out}^{4}}
\end{equation}
Therefore, the gravitational pressure over shell, $P_G$, is given
by
\begin{equation}
P_G = P_{in}-P_{out} =
\frac{-1}{960\pi^{2}}(\frac{1}{\alpha_{in}^{4}}-\frac{1}{\alpha_{out}^{4}}).
\end{equation}
Call the stress due to the boundary $P_B$. The total pressure on
the shell, $P$, is then given by
\begin{equation}
P = P_G + P_B =
\frac{-1}{960\pi^{2}}(\frac{1}{\alpha_{in}^{4}}-\frac{1}{\alpha_{in}^{4}})
+ \frac{\eta^{2}}{8\pi
a^{4}}(\frac{c_{1}}{\alpha_{in}^{2}}+\frac{c_{2}}{\alpha_{out}^{2}}).
\end{equation}
Noting the relation $\alpha^{2}=\frac{3}{\Lambda}$, we may write
the total pressure in terms of the cosmological constants:
\begin{equation}
P=\frac{-1}{2880\pi^{2}}(\Lambda_{in}^{2}-\Lambda_{out}^{2})
+\frac{\eta^{2}}{24\pi
a^{4}}(c_{1}\Lambda_{in}+c_{2}\Lambda_{out}).
\end{equation}
This total pressure may be both negative or positive. Note that
this is just the pressure due to quantum effects. Therefore the
following discussions should be taken cautiously. To see the
different possible cases, let us first assume
\begin{equation}
c_{1}\Lambda_{in}+c_{2}\Lambda_{out}>0,
\end{equation}
then $P_{B}>0$, i.e. the Casimir force on the bubble is repulsive.
Given a false vacuum inside and true vacuum outside, i.e.
$\Lambda_{in}>\Lambda_{out}$, then the gravitational part is
negative. Therefore the total pressure may be either negative or
positive. Given $P > 0$ initially, then the initial expansion of
the bubble leads to a change of the Casimir part of the pressure.
This change, depending on the detail of the dynamics of the
bubble, may be an increase or a decrease. Therefore, the initial
expansion of the bubble may end and a phase of contraction could
begin. For $P < 0$, there is an initial contraction which ends up
at a minimum radius. For the case of true vacuum inside and false
vacuum outside, i.e. $\Lambda_{in} < \Lambda_{out}$, which is
more interesting cosmologicaly, the total pressure is always
positive. Therefore the bubble expands
without any limit.\\
Now consider the case
\begin{equation}
c_{1}\Lambda_{in}+c_{2}\Lambda_{out} < 0.
\end{equation}
Noting that $|c_1| > |c_2|= - c_2$, it is seen that the inside has
to be a true vacuum, i.e. $\Lambda_{in} < \Lambda_{out}$.
Therefore, the total pressure may be either negative or positive.
For $P > 0$, the initial expansion of the bubble may be stopped
or not depending on the detail of the dynamics. For $P <0$, the
bubble contracts and the total pressure remains negative. Hence,
it ends up to a collapse of the bubble.

 \section{Conclusion}
  \vspace{3mm}
Spherical bubbles with different vacuums in- and out-side,
corresponding to different de Sitter space-times, are encountered
in infaltionary scenarios. To study the dynamics of such bubbles
one should know the Casimir effect on them. We have considered a
spherical shell in de Sitter background with a massless scalar
field, coupled conformally to the background, satisfying the
Dirichlet boundary conditions. Although the metric is time
dependent we could show that for a bubble with constant comoving
radius no particle is created. Our calculation shows that the
detail dynamics of the bubble depends on different parameters and
all cases of contraction, expansion and collapse may appear. The
interesting case of true vacuum inside leads to an expansion if
the difference of two vacuum energies is small. Otherwise the
bubble contracts and it leads to the collapse of the bubble.

\end{document}